\newcommand{\arctanh}{{\rm artanh} \,}
\newcommand{\tr}{{\rm tr} \,}
\documentstyle[sprocl,psfig]{article}

\def\be{\begin{equation}}
\def\ee{\end{equation}}
\def\bea{\begin{eqnarray}}
\def\eea{\end{eqnarray}}
\begin{document}

\title{Effective field theory for nuclear matter}

\author{Matthias Lutz}

\address{GSI, 64291 Darmstadt, Germany\\E-mail: m.lutz@gsi.de}


\maketitle\abstracts{We apply the relativistic chiral Lagrangian to the nuclear equation 
of state. An effective  chiral power expansion scheme, which is constructed to work around 
nuclear saturation density, is presented. The leading and subleading terms are 
evaluated and are shown to provide an equation of state with excellent saturation 
properties. Our saturation mechanism is found to probe details of the nuclear pion dynamics.}

\section{Introduction}

The nuclear equation of state awaits a fundamental theoretical description.
It is therefore important to develop the appropriate 
form of effective chiral perturbation theory (E$\chi $PT), a promising  
tool of modern nuclear physics, and apply it  to the 
nuclear many body problem systematically. 

The key element of any microscopic theory for the nuclear equation of 
state is the elementary nucleon-nucleon scattering process. In the 
context of chiral perturbation theory this problem was first addressed by Weinberg 
who proposed to derive a chiral nucleon-nucleon potential in time 
ordered perturbation theory \cite{Weinberg1,Bira}. 
In \cite{Lutz1,Lutz3} we proposed an alternative scheme with chiral power counting 
rules applied directly to the nucleon-nucleon 
scattering amplitude. This approach applies the manifest 
covariant version of the chiral Lagrangian and relies 
on the crucial observation that the chiral power counting rules can be 
generalized for 2-nucleon reducible diagrams. Non-perturbative effects like the 
pseudo bound state pole in the $^1S_0$ channel are generated by properly renormalized local 
two-nucleon vertices. A similar approach was suggested by Kaplan, Savage and Wise \cite{KSW} 
in the context of the chiral potential approach. Unfortunately the straightforward 
application of the extended counting rules to the s-wave nucleon-nucleon scattering phase 
shifts is hampered by a poor convergence and inconsistency with empirical 
data\cite{Cohen-Hansen}. In particular, unitarity is violated strongly even at small 
energy where one would expect this scheme to work well\cite{Lutz3}. This is, at first glance, 
a surprising result since it is long known that for the one pion 
exchange Yukawa potential perturbation theory works fine if the  
coupling strength is put to its empirical value\cite{Fogel,Friman,Steele}. 
In fact in the $^1S_0$ channel, the failure of a naive expansion is linked to the presence 
of a second strong subthreshold singularity reflecting a subtle interplay of intermediate 
attraction and short range repulsion\cite{Lutz3}. A properly modified unitary implementation 
of the extended chiral counting rules describes empirical s-wave elastic and inelastic 
nucleon-nucleon scattering in the spin singlet channel accurately up to 
$E_{\rm lab} \simeq 600 $ MeV\cite{Lutz3}.

We point out that counting rules for the vacuum nucleon-nucleon scattering amplitude  
greatly facilitate the construction of chiral power counting rules for the nuclear matter 
problem\cite{Lutz2}. Basically the density expansion or equivalently the multiple scattering 
expansion can be combined  smoothly with the chiral expansion where one identifies $k_F\sim Q$ 
as a further small scale. Such an identification leads to a systematic resummation of the 
density expansion avoiding the expansion in large ratios like $k_F/m_\pi$ but 
exploiting the chiral mass gap and expand in small ratios like 
$k_F/ m_\chi$ or $m_\pi/m_\chi $ with $m_\chi \simeq m_N$.

\section{Primer on relativistic chiral nucleon-nucleon scattering}

In this section we recall basic elements required for a successful  application of the 
relativistic chiral Lagrangian to the baryon sector. The systematic construction of the 
Lagrangian density in accordance with chiral symmetry was presented in great 
detail by Weinberg using stereographic coordinates \cite{Weinberg2}. There are two problems 
when evaluating the relativistic chiral Lagrangian in the baryon sector. First, any covariant 
derivative acting on the nucleon field produces the large nucleon mass and therefore must 
be assigned the minimal chiral power zero. 
Thus an infinite tower of interaction terms needs to be evaluated at given finite
chiral order \cite{Krause,Gasser}. Second, the straightforward evaluation of
relativistic diagrams involving nucleon propagators generates
positive powers of the large nucleon mass from loop momenta larger
than the nucleon mass. The chiral power counting rules are spoiled.
The relativistic scheme appeared impractical and the heavy
mass formulation of $\chi $PT, which overcomes both problems by
performing a non relativistic $1/m$ expansion at the level of the
Lagrangian density, was developed \cite{Manohar} and applied
extensively in the one-nucleon sector \cite{Bernard}.

We point out that in the two-nucleon channel it will be advantageous  to work with the 
relativistic form of the chiral Lagrangian \cite{Lutz1,Lutz3}. One reason is that in the 
heavy baryon formulation two-particle reducible diagrams are ill defined. A second reason 
is that a proper treatment of the pion production process $pp\rightarrow pp\,\pi^0$
requires relativistic kinematics\cite{Lutz3}. Note that an alternative scheme to tame 
potentially ill defined two-particle reducible diagrams is the use of time-ordered 
perturbation theory as suggested by Weinberg\cite{Weinberg1}. Since covariance is not 
manifest in time-ordered perturbation theory we prefer the relativistic scheme. 
Here we outline how the two aforementioned problems can be overcome within the 
framework of the relativistic chiral Lagrangian. Of course we expect the relativistic 
scheme to be equivalent to the more familiar heavy fermion formulation of chiral perturbation
theory in the one-nucleon sector. Rather than performing the $1/m$ expansion at the 
level of the Lagrangian density we suggest to work out this expansion explicitly
at the level of individual relativistic Feynman diagrams. Therewith we avoid the heavy 
baryon chiral Lagrangian with its known artifacts in the two-nucleon sector.
Here it is convenient to consider the chiral Lagrangian (at an intermediate stage) 
to represent a finite cutoff theory. The finite cutoff, $\Lambda \ll m$, is required to 
restrict the nucleon virtuality inside a given loop diagram such as to justify the $1/m$ 
expansion. 
Obviously a finite cutoff has to be introduced with great care not to break any chiral Ward
identity. Note that one may alternatively apply dimensional regularization. 
We emphasize that obviously if dimensional regularization is applied one has to {\it first} 
expand in $1/m$ and then perform the loop integration. The two steps do
{\it not} commute here. 

We need to outline how to arrange higher order terms of the 
relativistic chiral Lagrangian in the nucleon sector. To have a
practical scheme one would like to have only a finite number of
interaction terms with a given number of 'small' derivatives
involved. The known problem of the 'large' nucleon
mass generated by time derivatives acting on a nucleon field
has to be taken care of. This problem is cured easily as follows.
Consider a covariant derivative ${\mathcal D}_\mu (x) $ acting on a
nucleon field $N(x)$. We need to construct a 'counter interaction'
which leads to a small renormalized vertex. This rearrangement is
illustrated at hand of the simple case when the Lorentz index $\mu$
is saturated by another covariant derivative acting on the same 
nucleon field. In this case the counter interaction is readily found
\begin{eqnarray}
{\mathcal D}^\mu (x)\, {\mathcal D}_\mu(x) \, N(x) &&\rightarrow
\left({\mathcal D}^\mu (x)\, {\mathcal D}_\mu (x) +m^2 \right)\, N(x)
\nonumber\\
&&\rightarrow -\left(p^2-m^2 \right)
\sim Q
\label{eq8}
\end{eqnarray}
and the 'renormalized' vertex is proportional to the virtuality of
the nucleon 4-momentum which is of 'minimal' order $Q$. It is
important to observe that this counter interaction must have
occurred already in the chiral Lagrangian to some lower order since
it involves fewer derivatives. Note that a similar construction works if two 
covariant derivatives act on different nucleon fields.

We are left to consider the case in which the Lorentz index $\mu $
of the covariant derivative ${\mathcal D}_\mu $ is contracted
either with the Lorentz index of a Dirac matrix or the Lorentz
index of a derivative acting on a pion field. 
Interaction vertices in which a covariant nucleon
derivative ${\mathcal D}_\mu $ is saturated with a pionic four
vector $D^\mu $ or with the Lorentz index of its covariant
derivative behave well since they are already suppressed by the small derivative 
acting on the pion field. Here one has to assign the nucleon derivative the chiral
power $Q^{0}$. The large nucleon mass present is not hazardous
provided that one may identify the natural scale $\bar \Lambda$ of 
chiral interactions with the nucleon mass. With $\bar \Lambda \simeq m_\rho $ this 
seems indeed reasonable. The remaining case in which the nucleon derivative 
couples to the Lorentz index of a Dirac matrix does not generate an infinite tower of
interaction terms since at a given number of nucleon fields, $n_i$,
the number of available Lorentz indices provided by Dirac matrices
is at most $n_i$.

The above arguments show how to organize the infinite tower of
chiral interaction terms such that there is always only a finite
number of terms with a given number of 'small' derivatives. Thus at
any given chiral order only a finite number of diagrams need to be
evaluated. 
The result of this reorganization of the relativistic chiral Lagrangian leads to the 
applicability of Weinberg's counting rules provided that individual diagrams are 
expanded properly. 

In the two-nucleon sector the chiral Lagrangian leads to a well defined expansion scheme for 
the Bethe-Salpeter kernel K. For the unitary iteration of K, as induced by the Bethe-Salpeter 
equation, Weinberg's counting rules  are not applicable due to the s-channel unitarity cut.
Nevertheless a generalized counting for two-particle reducible diagrams can be established.
For a given diagram each pair of intermediate nucleons causes a reduction of one chiral power 
as compared to the 'naive' chiral power \cite{Lutz1}. The argument goes as follows: consider 
the once iterated Bethe-Salpeter kernel
\begin{eqnarray}
K\, G\, K
\sim
\int
\frac{d^4l }{(2\pi )^4} \frac{2\,m\,K(l)}
{\Big({\textstyle {1\over 2}}\,W-l\Big)^2-m^2+i\, \epsilon }
\frac{2\,m\,K(l)}{\Big({\textstyle {1\over 2}}\,W+l\Big)^2-m^2 +i\, \epsilon }
\label{bs2}
\end{eqnarray}
with $W^2 = s$. The object $G$ represents the two-nucleon propagator and $K$ the Bethe-Salpeter
kernel evaluated according to Weinberg's counting rules to a given chiral order. 
It is instructive first to study the s-channel spectral density, $\rho(s)$,
of this contribution. Since by assumption the Bethe-Salpeter kernel 
$K$ is 2-nucleon irreducible the s-channel spectral density picks up
strength only from the pinch singularity generated by the two
nucleon propagators in (\ref{bs2}). In the center of mass frame
with $\vec W =0 $ one finds:
\begin{eqnarray}
\rho(s) = \Im \,K\, G\, K \sim  \frac{p}{\sqrt{m^2+p^2}}\,\frac{m^2}{4 \, \pi }
\,\int \,
\frac{d \Omega_{\vec l} }{4\, \pi }\, K(\vec l\, )\, K(\vec l\,)\,
\Big|_{l_0 =0,\, |\vec l | = p }
\label{bs4}
\end{eqnarray}
with\footnote{Of course the iterated Bethe-Salpeter kernel does not necessarily
satisfy a s-channel dispersion relation. However, we point out that
only the part which provides strength for the s-channel spectral density
requires special attention and modified power counting rules. Obviously
a part which does not generate strength for the s-channel spectral
density has {\it no} pinch singularity leading to a s-channel cut so that
standard counting rules apply.} $s=4 \left( m^2 + p^2\right) $.
Expression (\ref{bs4}) can now be used to introduce the chiral power, $\nu$,
of the s-channel spectral density, $ \rho(s) $,
\begin{eqnarray}
\nu = 1+ 2\,\nu_K
\label{bs5}
\end{eqnarray}
with the chiral power $\nu_K$ of the Bethe-Salpeter kernel
$K$  as given by Weinberg's power counting rules. In (\ref{bs5}) we exploit the important 
observation that the small scale momentum $p\sim Q$ occurs in the spectral density.

To have a useful scheme we also need to derive counting rules for
the real part of the scattering amplitude. Here we evoke causality which relates 
the real part of a given Feynman diagram to its imaginary part 
by means of a dispersion relation
\begin{eqnarray}
I(s) = \Re K \, G\, K = \frac{1}{\pi}\,{\mathcal P } \int_{4\, m^2}^{\infty }d\,\bar s\,
\frac{\rho(\bar s)}{\bar s -s -i\,\epsilon} \; .
\label{bs7}
\end{eqnarray}
Of course we must not assume the dispersion relation (\ref{bs7}) to be
finite and well defined. It may be ultraviolet divergent\footnote{Note that for a given contribution one has to specify whether 
the dispersion relation holds at fixed Mandelstam variable $t$
or $u$. This technical detail, however, does
not affect our argument.}.
However, we recall that we are free to consider our chiral Lagrangian as
a finite cutoff theory. So let us introduce a covariant cutoff $\Lambda $
into the dispersion relation (\ref{bs7}). If the cutoff $\Lambda $ is
to be interpreted as a bound for the maximal virtuality of nucleons
one should first remove the nucleon rest mass. Therefore we take the cutoff $\Lambda $ 
to restrict the momentum $p\leq \Lambda $ rather than the
Mandelstam variable $s$:
\begin{eqnarray}
I(p, \Lambda ) = \frac{1}{\pi}\,{\mathcal P } \int_{0}^{\Lambda^2  }d\,\bar p^2\,
\frac{\rho(\bar p)}{\bar p^2 -p^2 } \; .
\label{bs8}
\end{eqnarray}
The natural guess for the chiral power of the integral $I(p)$ would be
the chiral power of its imaginary part. However, the integral
(\ref{bs8}) as it stands does not permit this chiral power yet. Though the
spectral density $\rho(p) $ exhibits only small scales as shown
above, the cutoff scale $\Lambda $ prevents the conclusion that the integral
behaves like a small scale to the power $1+2\,\nu_K $. We point out 
that the counting rule (13) can nevertheless be applied to the real part of 
the loop function, however, only after an appropriate  number of subtractions 
at $p^2=-\mu^2 < 0 $ with $\mu \sim m_\pi $. The subtraction is required to render 
the dispersion integral finite in the large cutoff limit: 
\begin{eqnarray}
I^{(n)}_S(p,\Lambda ;\mu ) =\frac{1}{\pi}\,{\mathcal P } \int_{0}^{\Lambda^2  }d\,\bar p^2\,
\frac{\rho(\bar p)}{\bar p^2 -p^2}
\left(\frac{p^2+\mu^2}{\bar p^2 +\mu^2}\right)^n \;.
\label{bs9}
\end{eqnarray}
The subtraction polynomial can always be absorbed into the local two-nucleon vertices.
Note that the performed subtractions do not necessarily
simplify our attempt to apply dimensional counting rules since this procedure
introduces a new scale, $\mu $, the subtraction point. Nevertheless
we now can exploit the freedom to choose the subtraction point. If we
insist on a 'small' subtraction scale $\mu \sim m_\pi $ of the
order of the pion mass one can apply dimensional power counting
since the integral is finite in the large cutoff limit by construction.
Thus we arrive at our desired chiral counting rule (13) for 2-particle reducible
diagrams: each pair of nucleon propagators which exhibits a s-channel
unitarity  cut gets a chiral power $-3$ in contrast to the 'naive'
power $-2$. We stress, first,
that this counting rule holds {\it only if} one introduces a 'small'
subtraction scale $\mu \sim m_\pi $ at the level of the Bethe-Salpeter
equation and second, that the subtraction scale is {\it independent}
of the intrinsic cutoff. We will refer to this counting rule
as the 'L'-counting rule since for the one pion exchange interaction the 
chiral power of the n-th iterated interaction is simply given by the number
of loops. In \cite{Lutz3} the reader may find explicit examples which confirm 
our L-counting rule (\ref{bs5}).  

The non-perturbative structures like the pseudo bound state in the spin singlet channel 
are generated naturally since the bare 2-nucleon vertex is renormalized strongly so that it 
effectively carries chiral power minus one. This can be seen as follows. 
Consider the leading order interaction
\begin{eqnarray}
{\mathcal L}_{4N}(x) &=&\frac{1}{4}\, g\,
\left(\bar N(x)\,\gamma_5 \,C\,\, \tau \tau_2\,\bar N^t(x)\right)
\left(N^t(x)\,\tau_2 \, \tau \, C^{-1}\gamma_5  \,N(x)\right) \; .
\label{mut0}
\end{eqnarray}
with the charge conjugation matrix $C= i\,\gamma_0\,\gamma_2 $ and the isospin 
Pauli matrices $\tau_i$. At tree level the coupling $g$ can be interpreted 
in terms of the s-wave scattering lengths $a(^1S_0) $ with
$2\,m\,g = -4\,\pi\, a(^1S_0)$. We assume here that the energy dependent 4-nucleon
bare coupling strengths, $g(p)$, is natural. According to our subtraction scheme, which  
is required for the manifestation of the L-counting rule, the bare coupling $g$ is 
renormalized by the one loop bubble $J$ at leading order:
\begin{equation}
g^{-1}(p^2)-J(p^2,\Lambda ) = g_R^{-1}(p^2,\mu,\Lambda ) - J_R(p,\mu ) \; .
\label{mut9}
\end{equation}
where the loop function $J$ is given by
\begin{eqnarray}
J(p^2,\Lambda )&=&-i\,\tr \int \frac {d^4 l}{(2\pi)^4} C^{-1}\gamma_5
\,S_F\left(l+{\textstyle {1\over2}} W \right)\gamma_5\, C
\,S_F\left(-l+{\textstyle {1\over2}} W \right) 
\nonumber\\
&=&\frac{1}{\pi} \int_{0}^{\Lambda^2 }
\frac {d\,\bar p^2\,\rho_J (\bar p )}{\bar p^2-p^2-i\,\epsilon}
\label{mut2}
\end{eqnarray}
in terms of the relativistic nucleon propagator $S_F(p)$ and $W^2=4\,(m^2+p^2)=s$. The renormalized
loop $J_R(p,\mu)=m\, (\mu+i\,p)/(2\,\pi)$ follows after the proper non-relativistic 
expansion of the spectral density:
\begin{eqnarray}
\rho_J (p )= \frac{m\,p }{2\pi}\,\sqrt{1+\frac{p^2}{m^2}}=
\frac{m\,p }{2\pi}\left(1 +\frac{p^2}{2\,m^2}+{\mathcal O}\left(\frac{p^4}{m^4}\right)\right)
\label{mut6} \; .
\end{eqnarray}
It is important to observe that for $p\leq \Lambda $ the renormalized coupling 
function can be expanded in powers of the momentum $p$ with
\begin{eqnarray}
g_R^{-1}(p^2,\Lambda ; \mu ) &=&  g^{-1}(p^2) -
\frac{m}{2\pi^2} \int_{0}^{\Lambda^2 }
\frac {d\,\bar p^2\,\bar p }{\bar p^2+\mu^2}\,
+\frac{m}{2\pi^2} \int_{\Lambda^2 }^{\infty }
\frac {d\,\bar p^2\,\bar p }{\bar p^2-p^2}\,
\frac{ p^2+\mu^2 }
     {\bar p^2+\mu^2}
\nonumber\\
&=&g^{-1}(p^2)-\frac{m}{\pi^2}
\left( \Lambda -\frac{\pi}{2}\, \mu  \right)
+\frac{m}{\pi^2} \frac{p^2}{\Lambda }\left(1 +{\mathcal
O}\left(\frac{p^2}{\Lambda^2}\right) \right)
\label{mut12}
\end{eqnarray}
where we dropped $1/m$ correction terms. Our result (\ref{mut12}) demonstrates that all cutoff 
dependence can safely be absorbed in the bare coupling $g(p)$ provided that it is legitimate 
to identify the typical cutoff $\Lambda $ with the natural scale of the bare coupling function.

Equation (\ref{mut12}) exhibits an interesting effect. Starting with a
perfectly natural coupling function $g(p^2)$ it potentially generates
an anomalously large renormalized coupling function
$g_R(p^2, \Lambda; \mu )$, however, only if the subtraction scale
$\mu \sim m_\pi $ is taken to be small. Then for an attractive
natural coupling $g<0 $ the $g^{-1}$ term is potentially canceled
by the $m\, \Lambda $ term such that the renormalized coupling $g_R$
will be anomalously large. We stress that this is a familiar phenomenon 
since it is long understood that an attractive system may dynamically
generate 'new' scales which are anomalously small (like for example the
binding energy of a shallow bound state). What is intriguing here is the transparent
and simple mechanism provided by (\ref{mut12}) to generate such a 'small'
scale naturally \cite{Lutz1,Lutz3,Beane}.

We close this section with a few simple examples demonstrating how to expand 
diagrams induced by the relativistic chiral Lagrangian and correct an erroneous 
statement made during my talk. It is sufficient to discuss 
scalar diagrams since any diagram with Dirac structure can always be
reduced to scalar loop function by simple algebra. Consider for example the scalar 
master loop functions which one naturally encounters in the evaluation of the relativistic 
pion nucleon box diagram:
\begin{eqnarray}
I_{3,N}(s)&=&i\,\int \frac{d^4l}{(2\pi)^4}\, S_\pi (l+{\textstyle
{1\over2}}\,Q)\, S_N(P-l)\,S_N(K+l)\; ,
\nonumber\\
I_{4,N}(s,t)&=&-i\,\int \frac{d^4l}{(2\pi)^4}\, S_\pi
(l+{\textstyle {1\over2}}\,Q)\, S_\pi (l-{\textstyle
{1\over2}}\,Q)\, S_N(P-l)\,S_N(K+l)
\label{}
\end{eqnarray}
with $S_i(p)=1/(p^2-m_i^2+i\,\epsilon )$. For technical convenience we introduce
the four vectors $P,K$ and $Q$ as follows:
\begin{eqnarray}
\begin{array}{ll}
p_1=K+{\textstyle {1\over2}} Q \; ,\;\;\;\;\;\;& p_1' = K-{\textstyle
{1\over2}} Q\; ,\\ p_2=P-{\textstyle {1\over2}} Q \; ,\;\;\;\;\;\;\;\;\;&
p_2' = P+{\textstyle {1\over2}} Q \; .
\end{array}
\end{eqnarray}
with $(P+K)^2=s$ and $Q^2=t$ but $P \cdot Q = K\cdot Q =0$. The master functions are easily 
calculated by means of dispersion techniques:
\begin{eqnarray}
I_{3,N}(s) &=&
\int_{4\, m^2}^{4\left( m^2+\Lambda^2\right)}
\frac{d\, s'}{\pi } \, \frac{\rho_{3,N }(s')}{s'-s}\; ,
\nonumber\\
I_{4,N}(s,t)&=&
\int_{4\, m^2}^{4\left( m^2+\Lambda^2\right)}
\frac{d\, s'}{\pi } \, \frac{\rho_{4,N}(s',t)}{s'-s} \;.
\label{disp-rep}
\end{eqnarray}
The finite cutoff $\Lambda < m $ is kept at this
intermediate stage (it will be removed later) in order to
mathematically justify the $1/m$ expansion. We derive the s-channel
spectral densities:
\begin{eqnarray}
\rho_{3,N}(s)&=&\frac{1}{16\,\pi }\frac{1}{\sqrt{4\, m^2 -u-t}}\,
\frac{1}{\sqrt{-u-t}}\,\ln
\left( \frac{m_\pi^2-u-t}{ m_\pi^2 }\right)\; ,
\nonumber\\
\rho_{4,N}(s,t) &=&\frac{1}{4\, \pi}\,\frac{1}{\sqrt{-t}}\,
\frac{1}{\sqrt{4\, m^2-u-t}}\, \frac{1}{\sqrt{b(s,t)}}\,
\arctanh \sqrt{\frac{t(u+t)}{b(s,t)}}
\label{}
\end{eqnarray}
with $u+t=4\,m^2-s$ and
\begin{eqnarray}
b(s,t) =4\,m_\pi^4
-4\,m_\pi^2 \left(u+t\right) + t \left( u+t\right)\; .
\label{}
\end{eqnarray}
The s-channel spectral densities are expanded in powers of the
inverse nucleon mass. At leading order we then derive:
\begin{eqnarray}
I_{3,N}(s) &=&\frac{1}{16 \pi m}
\,\frac{1}{\sqrt{-u-t}}\,
\Bigg(
\arctan \left(\frac{ \sqrt{-u-t}}{m_\pi}\right)
+\frac{i}{2}\,\ln
\left(1- \frac{u+t}{m_\pi^2 }\right)\Bigg)\; ,
\nonumber\\
I_{4,N}(s,t) &=&
\frac{1}{8\,\pi m}\,\frac{1}{\sqrt{-t\,b(s,t)}}
\Bigg( i\,\arctanh  \left(\sqrt{\frac{t(u+t)}{b(s,t)}}\,\right)
\nonumber\\
&+&\frac{1}{2}\,\arctan
\left(  \frac{\sqrt{-t\,b(s,t) }}{4\,m_\pi^3}
-\frac{\sqrt{-u-t}}{m_\pi } \left(1-\frac {t}{4\,m_\pi^2} \right) \right)
\nonumber\\
&+&\frac{1}{2}\,\arctan
\left(  \frac{\sqrt{-t\,b(s,t) }}{4\,m_\pi^3}
+\frac{\sqrt{-u-t}}{m_\pi } \left(1-\frac {t}{4\,m_\pi^2} \right) \right)
\Bigg)
\label{master-loops-results}
\end{eqnarray}
where we expand in $\Lambda/m $, $u/m$ and $t/m$. Note that the $(1/m)^n$ correction terms of
$I_{3,N}(s)$ and $I_{4,N}(s,t)$ can be evaluated in closed form (up to a scheme dependent 
subtraction polynomial) by simply replacing the $1/m$ factor in (\ref{master-loops-results}) by the factor 
$2/\sqrt{s}$. This is an immediate consequence of the representation (\ref{disp-rep}). We find
this example highly instructive since it clearly demonstrates how the difference of the 
full relativistic loop function and the $1/m$ expanded loop is lumped efficiently into the
subtraction polynomial. 

It is instructive to confront our relativistic scheme with the potential approach. 
Inherent of the potential approach is the use of the static pion exchange. 
The pion propagator is expanded
\begin{eqnarray}
S_\pi(l_0,\vec l\,) &=& \frac{-1}{ l^2+m_\pi^2} -\frac{l^2_0}{(l^2+m_\pi^2)^2}+
{\mathcal O} \left(l_0^4 \right)
\label{pi-static}
\end{eqnarray}
where it is commonly argued that terms involving the energy transfer $l^2_0 $ are 
suppressed by $1/m $. Mathematically it is not quite obvious that such an expansion is 
justified and in accordance with covariance since the expansion is highly divergent due to 
multiple powers of the energy transfer $l_0$. We evaluate the master pion nucleon box 
function $I_{4,N}(u,t)$ in the limit of static pions
\begin{eqnarray}
I^{(static )}_{4,N} (\vec p,\vec p\,' ) &=& \frac{1}{4\,m}
\int \frac{d^3l}{(2\pi)^3}\,\frac{1}{(\vec l+\vec p\,)^2+m_\pi^2}
\,\frac{1}{(\vec l+\vec p\,'\,)^2+m_\pi^2} \,\frac{1}{l^2-p^2-i\,\epsilon }
\nonumber\\
\label{static-result}
\end{eqnarray}
where we chose the center of mass frame with $p^2=p'^2$ and 
$\vec p\cdot \vec p' = p^2\,\cos \theta $ and $s=4\,(m^2+p^2)$. Note that we dropped terms 
suppressed by $1/m^3$. Explicit evaluation of the integral (\ref{static-result}) leads in 
fact to the same expression derived before in (\ref{master-loops-results}). It remains to
be shown  that also the $(1/m)^n$ correction terms evaluated in such a scheme agree with 
those of the relativistic approach. Note that in any case it becomes more and more tedious to 
evaluate the $(1/m)^n$ correction terms using (\ref{pi-static}).

In the algebraic reduction of the pion nucleon box diagram one 
encounters a further master loop function
\begin{eqnarray}
I_{3,\pi }(t) &=&i\,\int
\frac{d^4l}{(2\pi)^4}\, S_\pi (l+{\textstyle {1\over2}}\,Q)\, S_\pi
( l-{\textstyle {1\over2}}\,Q)\, S_N(l+K) 
\label{}
\end{eqnarray}
also relevant for the pion-nucleon scattering process. 
The triangle loop, $I_{3,\pi }(t)$, is finite and can be evaluated by means of a
dispersion integral:
\begin{eqnarray}
I_{3,\pi }(t) &=&\int_{4\, m_\pi^2}^{4\left(
m_\pi^2+\Lambda^2\right)}
\frac{d\, t'}{\pi } \, \frac{\rho_{3,\pi }(t')}{t'-t}
\label{i3-disp}
\end{eqnarray}
with the t-channel spectral density:
\begin{eqnarray}
\rho_{3,\pi }(t)&=&\frac{1}{8\, \pi }\, \frac{\Theta \Big( t-4\,m_\pi^2\Big)}
{\sqrt{t(4\, m^2-t)}}\,
\arctan \left( \sqrt{4\, m^2-t}\,
\frac{\sqrt{t-4\, m_\pi^2}}{t-2\, m_\pi^2}\right)
\label{i3-pi-spec}
\end{eqnarray}
for $t>4\, m_\pi^2 $. Naively one may consider the spectral density
in the heavy nucleon mass limit with
\begin{eqnarray}
\rho_{3,\pi}(t)=\frac{1}{32\,m\,\sqrt{t}}\,\Theta \Big( t-4\,m_\pi^2\Big)
\label{}
\end{eqnarray}
and derive
\begin{eqnarray}
I_{3,\pi }(t) =\frac{1}{16\,\pi\,m}\frac{1}{\sqrt{-t}}\,
\arctan \left(\frac{\sqrt{-t}}{2\,m_\pi} \right) 
\label{i3pi-lead}
\end{eqnarray}
for $t<0$. We point out, however, that the $1/m$ expansion of the loop integral 
$I_{3,\pi }(t) $ is subtle. This can be anticipated from the fact that the $1/m$ 
expansion of the t-channel spectral density leads, at subleading orders, to singularities 
in the dispersion integral at $t'=4\,m_\pi^2 $\footnote{Note that such singularities 
disappear in the chiral limit with $m_\pi=0$.}. Note that the 'naive' leading 
order result (\ref{i3pi-lead}) is in agreement with the corresponding expression derived 
in the heavy baryon framework \cite{Tang,Mojzis}. 
An alternative way to perform the non-relativistic expansion is through the exact identity
\begin{eqnarray}
I_{3,\pi}(t) =I_{3,N}(4\,m^2-t)
-\left( m_\pi^2-{\textstyle {1\over2}}\,t\right) I_{4,N}(4\,m^2-t,t )
\label{id-exact}
\end{eqnarray}
which holds for the full relativistic loop functions. It expresses the desired loop
integral $ I_{3,\pi}(t)$, in terms of the master loop functions
$I_{3,N}(s)$ and $I_{4,N}(s,t)$ which posses a well defined $1/m$
expansion. The non-relativistic expansion of $I_{3,N}(s)$ and
$I_{4,N}(s,t)$ then confirms (\ref{i3pi-lead}). We suggest that the proper $(1/m)^n$ 
correction terms of $I_{3,\pi}(t) $ are in fact accessed via (\ref{id-exact}). In terms 
of the t-channel dispersion relation (\ref{i3-disp}) this  
amounts  to only  including the $1/m$ correction terms from the factor $
1/\sqrt{4\,m^2-t}$  but use the large nucleon mass limit of the 
$\arctan $-term in (\ref{i3-pi-spec}).

The few examples presented here clearly demonstrate the usefulness of the  relativistic
chiral Lagrangian. The required $1/m$ expansion of Feynman diagrams
is straightforward and in fact performed most economically applying dispersion 
relation techniques introduced by Cutkosky \cite{Cutkosky}. These techniques keep covariance 
manifestly and are applied easily also to higher loop diagrams involving pion 
production cuts\cite{Lutz2}. Note that dealing with relativistic Feynman diagrams avoids 
the cumbersome splitting of pions into static, soft and radiation pions as suggested by 
Mehen and Stewart\cite{Mehen2}.

\section{Chiral expansion scheme for nuclear matter}

An attempt to apply the generalized chiral power counting rules of 
\cite{Lutz1,Lutz3} to the nuclear many body problem quickly reveals that 
even though the pion dynamics remains perturbative the local 
two-nucleon interaction requires extensive summations. This is an immediate 
consequence of the chiral power given to the renormalized coupling $g_R\sim Q^{-1}$.
In nuclear matter at small density with $k_F \simeq 100$ MeV and $\rho=2\,k_F^3/(3\,\pi^2)$
it is not sufficient to sum the particle-particle ladder diagrams of the local two-nucleon 
interaction. 
We find that one must sum simultaneously also the particle-particle and hole-hole ladder 
diagrams including all interference terms where a self consistently dressed nucleon 
propagator must be used. According to the generalized counting rules the pions can 
then be evaluated perturbatively on top of the above described parquet resummation for  
the local two-nucleon interaction\footnote{Our expectation that 
the nuclear equation of state can be calculated microscopically solving the parquet theory 
for the local interaction and then include pions perturbatively may be somewhat too naive. 
A refined formulation which incorporates relevant subthreshold singularities of the
vacuum nucleon-nucleon scattering amplitude (see \cite{Lutz3}) may lead to a more involved 
treatment of pionic effects.}. 
In terms of vacuum scales this leads to an expansion of the energy per particle, $\bar E(k_F)$, of the 
form 
\begin{eqnarray}
\bar E(k_F) &=& \sum_n\, \bar E_n\!\!\left(\frac{k_F}{m_\pi}, 
\frac{k_F}{\Lambda_S}\right)  \left( \frac{k_F}{\Lambda_L}\right)^n
\label{exp2}
\end{eqnarray}
with typical small scales, $\Lambda_S$, like  $ \sqrt{m_N 
\,\epsilon_D} $ where $\epsilon_D \simeq 2$ MeV is the deuteron binding energy, and 
typical large scales $\Lambda_L \simeq 4\,\pi \,f_\pi \simeq 1$ GeV. The expansion 
coefficients $\bar E_n$ are 
complicated and hitherto unknown functions of the Fermi momentum $k_F$. 
They can be computed in terms of the free space chiral Lagrangian furnished with a 
systematic summation technique \cite{Lutz:prep}. 

In this work we pursue a somewhat less microscopic approach in 
spirit close to the Brueckner scheme but more systematic in the 
sense of effective field theory. Since  the typical small scale 
$\Lambda_S $ is much smaller than the Fermi momentum $k_{F,0} 
\simeq 265 $ MeV at nuclear saturation density, one may expand the 
coefficient functions $\bar E_n$ around $k_{F,0} $ in the 
following manner 
\begin{eqnarray}
\bar E_n\!\!\left(\frac{k_F}{m_\pi}, 
\frac{k_F}{\Lambda_S}\right) &=& \bar E_n\!\!\left(\frac{k_F}{m_\pi}, 
\frac{k_{F,0}}{\Lambda_S}\right)
+\sum_{k=1}^\infty \bar E_n^{(k)}\!\!\left(\frac{k_F}{m_\pi} 
\right) 
\left( \frac{\Lambda_S }{k_F}-\frac{\Lambda_S}{k_{F,0}}\right)^k
\; .
\label{exp3}
\end{eqnarray}
Note that we do not expand in the ratio $m_\pi/k_F $. If one 
expanded also in this ratio $m_\pi/k_F $ one would arrive at the 
Skyrme phenomenology \cite{Skyrme} applied successfully to nuclear 
physics many years ago. It should be clear that this scheme is 
constructed to work around nuclear saturation density but will fail 
at small density. We note that also conventional approaches like 
the Walecka mean field or the Brueckner scheme are known to be 
incorrect at small density. 

In terms of diagrams  one may arrive at the proposed 
expansion (\ref{exp3}) by a perturbative evaluation of the pionic nuclear 
many-body effects on top of the the high density limit of the parquet 
resummation for the local two-nucleon interaction. The underlying assumption is of course
that the typical small scale in the parquet, $\Lambda_S$, is such that the high density limit 
is reached already at densities much smaller than the nuclear saturation density $k_{F,0}$.
Technically our scheme can be generated by the effective Lagrangian density 
\begin{eqnarray}
{\mathcal L}_{int}(k_F) &=& \frac{g_A}{2\,f_\pi} \, \bar N
\gamma_5\, \gamma^\mu \cdot 
\left( \partial_\mu  \,{\vec \pi }\right) \cdot {\vec \tau} \,N
\nonumber\\
&+& \frac{1}{8\,f_\pi^2}\,\left(g_0(k_F)+\frac{1}{4}\,g_A^2 \right)
\left( \bar N \,\gamma_5\,\tau_2\, C^{-1}\,\bar N^t\right)
\left( N^t\, C\,\tau_2\,\gamma_5\,N \right)
\nonumber\\
&+& \frac{1}{8\,f_\pi^2}\,\left(g_1(k_F)+\frac{1}{4}\,g_A^2 \right)
\left( \bar N \,\gamma_\mu\,\vec \tau \,\tau_2\, C^{-1}\,\bar N^t\right)
\left( N^t\, C\,\tau_2\,\vec \tau \,\gamma^\mu\,N \right)
\label{l1}
\end{eqnarray}
where the couplings $g_0=g_0(k_F)$ and $g_1=g_1(k_F)$ are density 
dependent. A more systematic derivation of the expansion 
(\ref{exp2}) and (\ref{exp3}) applying suitable resummation 
techniques will be presented elsewhere \cite{Lutz:prep}. 

\begin{figure}[t]
%\rule{5cm}{0.2mm}\hfill\rule{5cm}{0.2mm}
%\vskip 2.5cm
%\rule{5cm}{0.2mm}\hfill\rule{5cm}{0.2mm}
\begin{center}
\psfig{figure=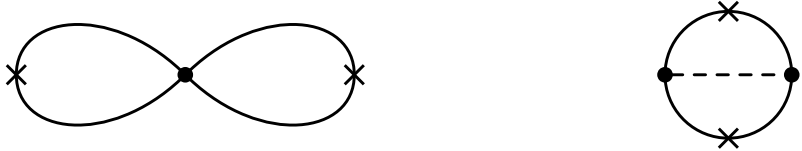,height=0.65in}
\end{center}
\caption{Leading contribution of chiral order $Q^6$.  \label{fig2}}
\end{figure}

The chiral power counting rules are simplified significantly as 
compared to a fully microscopic scheme. The presence of a further 
small scale $k_F \sim Q \sim m_\pi $ does not anymore generate an 
infinite tower of diagrams, to be considered at a given chiral 
order, since by construction the troublesome local 2-nucleon vertex 
need not be iterated, i.e. terms proportional to $g_0^n(k_F)$ and 
$g_1^n(k_F)$ with $n>1$ are already included in $g_0(k_F)$ and 
$g_1(k_F)$ and therefore must not be considered. Note that here we count
$g_{0,1} \sim Q^0$ since the non-perturbative structures like the deuteron, which 
give rise to the anomalous power $g_{0,1}\sim Q^{-1}$ in the vacuum,  
are long dissolved at densities close to nuclear saturation.
The pion dynamics, if properly renormalized, remains perturbative like in the vacuum 
case. Consider for example the two loop diagrams depicted in Fig. 
\ref{fig2} where the nucleon line with a 'cross' represents the projector onto 
the Fermi sphere 
\begin{eqnarray}
\Delta S_N(p) = \left( \gamma \cdot p+m_N \right)
2\,\pi\,i\,\Theta(p_0)\,\delta (p^2-m_N^2)\,
\Theta \left(k_F^2-\vec p\,^2\right)\; .
\label{cross}
\end{eqnarray}
The first diagram in Fig. \ref{fig2} is proportional to 
$g_{0,1}(k_F) \, k_F^6$ and is therefore ascribed the chiral order 
$Q^6 $ since the effective vertex $g_{0,1}(k_F) \sim Q^0 $ carries 
chiral power zero. The second diagram, the one pion exchange 
contribution, is also of chiral order $Q^6$ since it is 
proportional to $k_F^6 $ multiplied with some dimensionless 
function $f(m_\pi/k_F)$. 
\begin{figure}[t]
%\rule{5cm}{0.2mm}\hfill\rule{5cm}{0.2mm}
%\vskip 2.5cm
%\rule{5cm}{0.2mm}\hfill\rule{5cm}{0.2mm}
\begin{center}
\psfig{figure=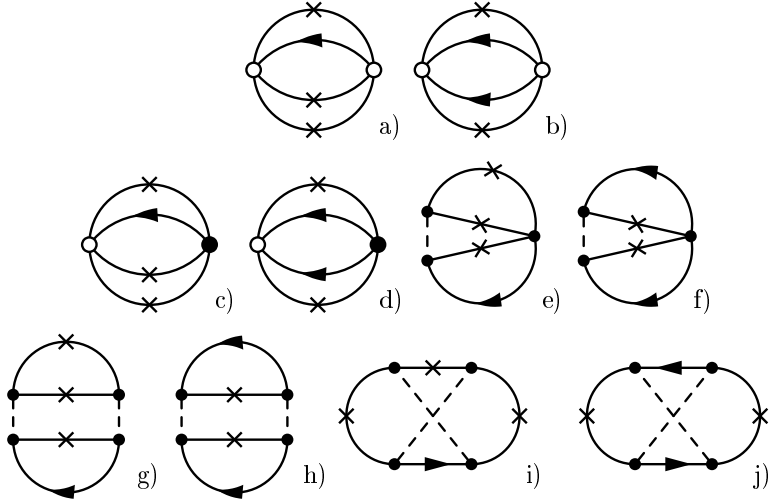,height=2.85in}
\end{center}
\caption{Subleading contribution of chiral order $Q^7$.  \label{fig3}}
\end{figure}
In Fig. \ref{fig3} we collected all diagrams of chiral order $Q^7$. 
Here we introduced two types of 2-nucleon vertices. The filled 
circle represents the full vertex of (\ref{l1}) proportional  to 
$g_{0,1}+g_A^2/4 $ and the open circle the counter term 
proportional to $g_A^2/4 $. The dashed line is the pion propagator 
and the directed solid line the free space nucleon propagator. We 
point out that the diagrams b), d), f), h) and j) in Fig. \ref{fig3} 
are divergent. The leading chiral contribution of the sum of all 
diagrams, however, is finite after including the appropriate counter term 
in the spin triplet channel.
\begin{figure}[t]
%\rule{5cm}{0.2mm}\hfill\rule{5cm}{0.2mm}
%\vskip 2.5cm
%\rule{5cm}{0.2mm}\hfill\rule{5cm}{0.2mm}
\psfig{figure=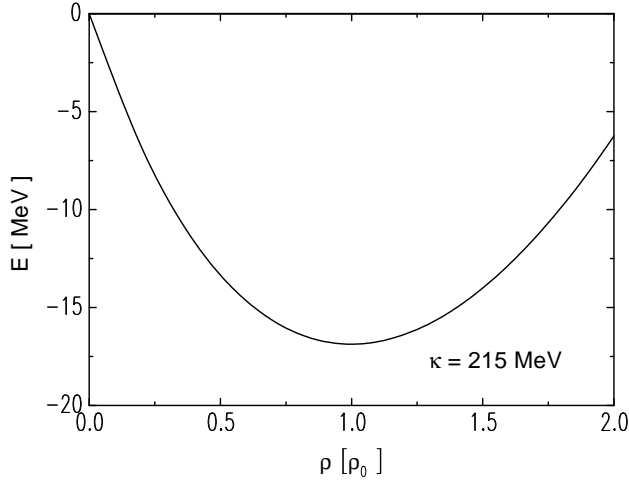,height=2.5in}
\caption{The equation of state for isospin symmetric nuclear matter.  \label{fig5}}
\end{figure}

It is instructive also to display the non relativistic form of the 
Lagrangian density (\ref{l1}) as suggested by Weinberg 
\cite{Weinberg1} in the context of the nucleon-nucleon scattering 
problem  
\begin{eqnarray}
{\mathcal L}_{int}(k_F) &=&\frac{g_A}{2\,f_\pi} \, \bar N
\left(\vec \sigma \cdot \vec \nabla \right)\Big(
{\vec \pi } \cdot {\vec \tau} \Big)\,N
\nonumber\\
&+& \frac{1}{8\,f_\pi^2}\,\left(g_0(k_F)+\frac{1}{4}\,g_A^2 \right)
\left( \bar N \,\vec \sigma\,\sigma_2\,\tau_2\,\bar N^t\right)
\left( N^t\,\tau_2\,\sigma_2\,\vec \sigma \,N \right)
\nonumber\\
&+& \frac{1}{8\,f_\pi^2}\,\left(g_1(k_F)+\frac{1}{4}\,g_A^2 \right)
\left( \bar N \,\sigma_2\,\,\vec \tau \,\tau_2\,\bar N^t\right)
\left( N^t\, \tau_2\,\vec \tau \,\sigma_2\,N \right)
\label{l2}
\end{eqnarray}
with $N$ now a two component spinor field and $\bar N =N^\dagger $ 
for notational convenience. According to Weinberg the two-particle 
reducible diagrams are to be evaluated with static pions.

We therefore evaluate all diagrams of 
Fig. \ref{fig3} with the appropriate non relativistic 
interaction vertices and static pion propagators. The 
solid line with a 'cross' now represents the non relativistic limit 
of (\ref{cross}) and the directed solid line the free non 
relativistic nucleon propagator. For technical details we refer to 
\cite{Lutz:prep2}. Both schemes indeed lead to identical results for 
all diagrams\footnote{Note that here we correct an error in \cite{Lutz2}.}.

In Fig. \ref{fig5} our result for the isospin symmetric nuclear 
equation of state is shown. The relevant coupling $g_0+g_1$ is adjusted to obtain nuclear 
saturation at $k_{F,0} = 265 $ MeV. We 
emphasize that the coupling functions $g_{0,1}(k_F)$  are to be 
determined from the nuclear equation of state.  However there is a 
strong consistency constraint: according to our 
scale argument (\ref{exp3}) the density dependence of the couplings 
$g_{0,1}(k_F)$ must be weak for $k_F$ larger than typical small scales integrated 
out. If the nuclear saturation required a 
strong density dependence our scheme had to be rejected. 
The density independent set of parameters $g_0+g_1 
\simeq 3.2 $, $g_A \simeq 1.26 $, $m_\pi \simeq 140 $ MeV  and 
$f_\pi \simeq 93 $ MeV give an excellent result for the equation of 
state. The empirical saturation density $k_{F,0} \simeq 265 $ MeV 
and the empirical binding energy of 16 MeV are reproduced. The 
incompressibility with $\kappa \simeq 215 $ MeV is also compatible 
with the empirical value $(210 \pm 30) $ MeV of \cite{Blaizot}. 

\section{The chiral order parameter $\langle \bar qq \rangle $}

The quark condensate, $\langle \bar qq \rangle (\rho )$, is an 
object of utmost interest. It measures the degree of chiral 
symmetry restoration in nuclear matter. Furthermore it is an 
important input for QCD sum rules \cite{S.H.Lee} or the Brown-Rho 
scaling hypothesis \cite{BR-scaling}. According to the Feynman-Hellman 
theorem the quark condensate can be extracted unambiguously 
from the total energy, $E(\rho) $, of nuclear matter once the 
current quark mass dependence of $E(\rho , m_Q )$ is known 
\begin{eqnarray}
\langle \bar qq \rangle (\rho ) &=& \frac{1}{V}\,
\frac{\partial}{\partial\,m_Q} \, E(\rho, m_Q )
\label{FH1}
\end{eqnarray}
with $m_Q=m_u=m_d$. Recall that $ E(\rho )/V 
= ( m_N+\bar E(\rho ) )\,\rho $ is determined by the nuclear 
equation of state $\bar E(\rho ) $. Since our chiral approach was 
set up to treat the pion dynamics and therewith the current quark 
mass dependence of the equation of state properly it is very much 
tailored to be applied to the quark condensate.  

\begin{figure}[t]
%\rule{5cm}{0.2mm}\hfill\rule{5cm}{0.2mm}
%\vskip 2.5cm
%\rule{5cm}{0.2mm}\hfill\rule{5cm}{0.2mm}
\psfig{figure=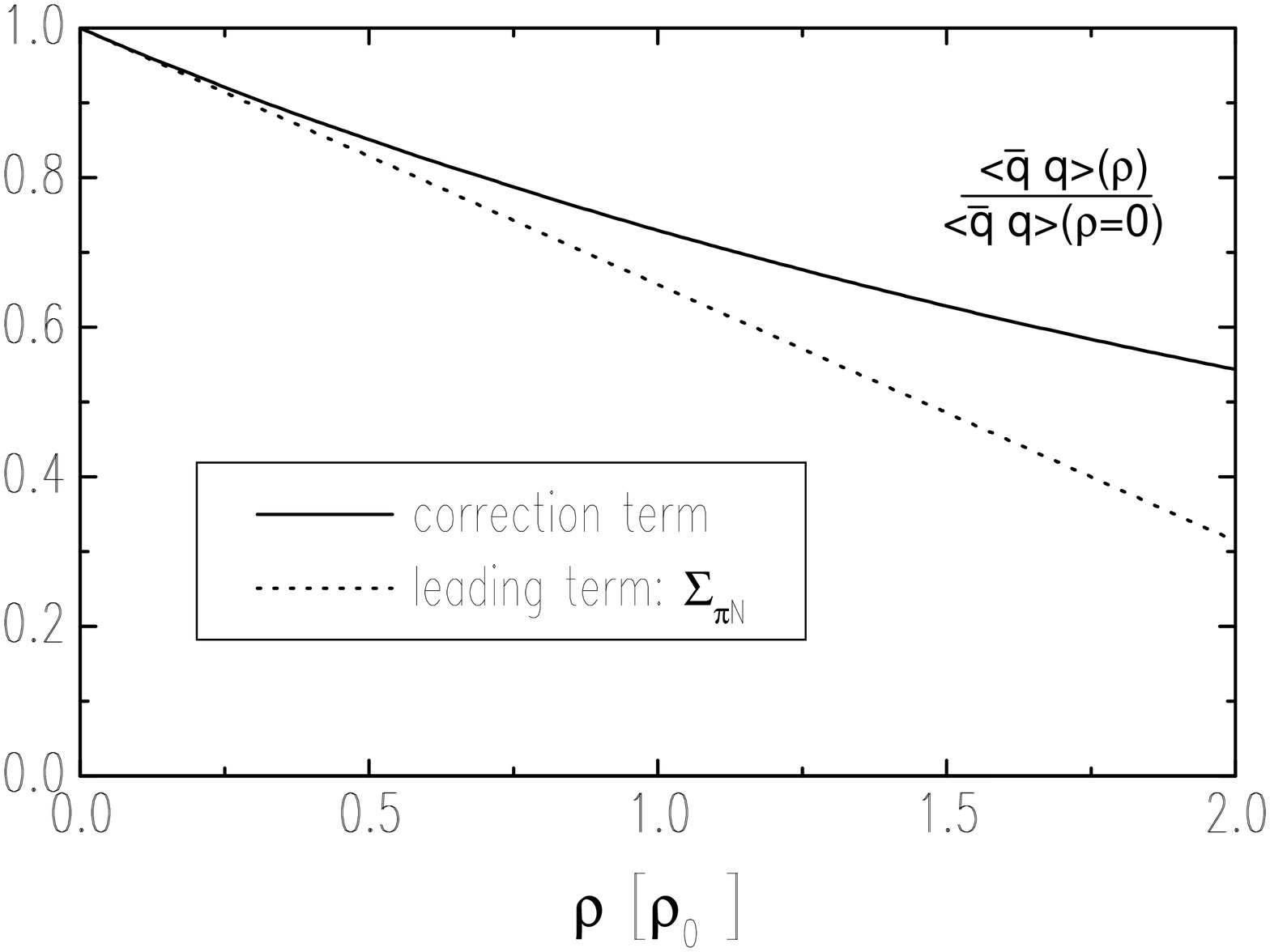,height=2.5in}
\caption{The quark condensate in isospin symmetric nuclear matter.  \label{fig6}}
\end{figure}

It is convenient to consider the relative change of the quark 
condensate since it is  renormalization group invariant: 
\begin{eqnarray}
\frac{\langle \bar q\, q \rangle (\rho)}
{\langle \bar q\, q \rangle (0)}
=1-\frac{\Sigma_{\pi N}\,\rho}{m_\pi^2\,f_\pi^2}
-\frac{\alpha_\pi(\rho)\,\rho}{2\,m_\pi\,f_\pi^2}
\label{FH2} \; .
\end{eqnarray}
The second term in (\ref{FH2}) follows from the nucleon rest mass 
contribution to the total energy of nuclear matter together with 
the definition of the pion nucleon sigma term, $\Sigma_{\pi N} $: 
\begin{eqnarray}
\Sigma_{\pi N} = m_Q \,\langle N|\bar q\, q |N\rangle
=m_Q\,\frac{d\,m_N}{d\,m_Q} \; .
\label{}
\end{eqnarray}
The last term in (\ref{FH2}) measures the sensitivity of the 
nuclear equation of state, $\bar E(\rho, m_\pi )$, on the pion mass 
\begin{eqnarray}
\alpha_\pi(\rho) &=& 
-\frac{2\,m_\pi\,f_\pi^2}{\langle \bar q\, q \rangle (0)}
\,\frac{\partial }{\partial\, m_Q}\,\bar E(\rho ,m_\pi )
=\left(1+{\mathcal 
O}\left( m_\pi^2 \right) \right)\frac{\partial }{\partial 
\,m_\pi}\,\bar E(\rho, m_\pi ) 
\label{}
\end{eqnarray}
where we consider now the current quark mass $m_Q=m_Q(m_\pi)$ as a 
function of the physical pion mass. We emphasize that the second 
term in (\ref{FH2}) written down first in 
\cite{Drukarev,Lutz92,Cohen} does not probe the nuclear many body 
problem and therefore should be considered with great caution. It 
is far from obvious that this term is the most important one at 
nuclear saturation density. The only term susceptible to nuclear 
dynamics is the last term in (\ref{FH2}).

In Fig. \ref{fig6} we present our result for the quark condensate 
in nuclear matter. We confront the 'leading' term driven by 
$\Sigma_{\pi N}\simeq 45 $ MeV with  our result.  
The inclusion of pionic many body effects leads to a significantly less reduced quark 
condensate in nuclear matter. Note that here we do not include an explicit 
$m_\pi$ dependence of the coupling $g_{0,1}(k_F)$ simply because there is not yet any 
reliable estimate available.  
Our results confirm calculations performed within 
the Brueckner \cite{Li} and Dirac-Brueckner \cite{Brockmann} approach qualitatively insofar 
that the nuclear many body system appears to react against chiral symmetry restoration.  
They have strong implications for the QCD sum rule 
approach of hadron properties in nuclear matter.

\end{document}